\documentclass[%
preprint,
showpacs,preprintnumbers,
nobibnotes,
 amsmath,amssymb,
 aps,
 pre,
 longbibliography,
 lengthcheck,%
]{revtex4-1}

\usepackage{graphicx}% Include figure files
\usepackage{dcolumn}% Align table columns on decimal point
\usepackage{bm}% bold math
\usepackage{hyperref}% add hypertext capabilities

\begin{document}

\title{Noise-induced rectification in out-of-equilibrium structures}

\author{R. Salgado-Garc\'{\i}a} 
\email{raulsg@uaem.mx}
\affiliation{Centro de Investigaci\'on en Ciencias-IICBA, Universidad Aut\'onoma del Estado de Morelos. Avenida Universidad 1001, Colonia Chamilpa, 62209, Cuernavaca Morelos, Mexico.} 
\date{\today} 

\begin{abstract}

We consider the motion of overdamped particles over random potentials subjected to a Gaussian white noise and a time-dependent periodic external forcing. The random potential is modeled as the potential resulting from the interaction of a point particle with a random polymer. The random polymer is made up, by means of some stochastic process,  from a finite set of possible monomer types.  The process is assumed to reach a non-equilibrium stationary state, which means that every realization of a random polymer can be considered as an out-of-equilibrium structure. We show that the net flux of particles over this random medium is non-vanishing when the potential profile on every monomer is symmetric. We prove that this ratchet-like phenomenon is a consequence of the irreversibility of the stochastic process generating the polymer. On the contrary, when the process generating the polymer is at equilibrium (thus fulfilling the detailed balance condition) the system is unable to rectify the motion. We calculate the net flux of the particles in the adiabatic limit for a simple model and we test our theoretical predictions by means of Langevin dynamics simulations. We also show that, out of the adiabatic limit, the system also exhibits current reversals as well as non-monotonic dependence of the diffusion coefficient as a function of forcing amplitude.

\end{abstract}

%\pacs{05.40.-a,05.10.Gg,05.70.Ln}

\pacs{05.40.−a,05.60.−k,05.10.Gg,02.50.Cw}

%% Creating title
\maketitle

%$$$$$$$$$$$$$$$$$$$$$$$$$$$$$$$$$$$$$$$$$$$$$$$$$$$$$$$$$$$$$$
%$$$$$$$$$$$$$$$$$$$$$$$$$$$$$$$$$$$$$$$$$$$$$$$$$$$$$$$$$$$$$$
\section{Introduction}

%$$$$$$$$$$$$$$$$$$$$$$$$$$$$$$$$$$$$$$$$$$$$$$$$$$$$$$$$$$$$$$
%$$$$$$$$$$$$$$$$$$$$$$$$$$$$$$$$$$$$$$$$$$$$$$$$$$$$$$$$$$$$$$

Several structures in nature are known to be arisen under non-equilibrium conditions. Formation of glasses is one archetypical example of this situation, since a glass can be viewed as a liquid that has lost its ability to flow~\cite{Angell1995}. Another paradigm of out-of-equilibrium structures is the DNA molecule. Although many of the statistical features of the genome are not well understood, it is commonly accepted that the DNA is a structure having some characteristics of systems out of equilibrium~\cite{Provata2014,rsg2016symbolic}. For example, it has been shown that warm-blooded vertebrates has a ``mosaic organization''  with respect to the variation of the GC content along the genome~\cite{bernardi1989isochore}. Another characteristic is, for instance, that the DNA has a strong deterministic component in its structure, which is a consequence of the fact that the topological entropy is practically  zero for blocks longer than 12 base-pairs~\cite{rsg2016symbolic}. Moreover, in a recent work~\cite{Provata2014} Provata \emph{et al} have tackled directly the problem of determining if real DNA has some statistical characteristics of non-equilibrium structures. Particularly they found that the detailed balance does not hold for human DNA, suggesting that the genome is spatially asymmetric and irreversible. 

Here we are interested in the dynamics of a Brownian particle when it moves on an out-of-equilibrium structure. This study is motivated by the fact that there are some proteins that slide along DNA, a process which is of importance in many biological functions~\cite{Kolomeisky2011physics,Bressloff2011stocastic,Cocho2003replication}. If DNA can be considered as a non-equilibrium structure, a natural question that raises from this observation is if this property affects in some way the transport properties of particles moving on DNA, such as the particle current or the diffusion coefficient. In this work we study the influence of non-equilibrium features of a medium on the transport properties of Brownian particles. To achieve this goal we model the substrate as a ``random polymer'' produced by a simple stationary Markov process out of equilibrium. This model gives rise to ``polymers'' having a spatial irreversibility due to the fact that the detailed balance does not hold. Then we use the particle-polymer model that has been studied in Refs.~\cite{rsg2013normal,rsg2014effective,rsg2015unbiased,rsg2016normal,hidalgo2017scarce} to study the dynamics of Brownian particles on random potentials. Particularly we show that, under the influence of an unbiased time-dependent periodic forcing, the spatial irreversibility of the medium induces a rectification phenomenon similar to the one occurring in the so-called rocked thermal ratchets. The rectification phenomenon we report is induced by the interplay between the thermal noise, the external forcing and the spatial irreversibility of the substrate. It is worth emphasize that the ``rectification phenomenon'' that we report here does not arise from an asymmetric potential profile (a necessary characteristic for ratchet systems to operate), making this rectification mechanism different from the one of ratchet systems. Besides the rectification phenomenon, we observe in our model other transport properties, out of the adiabatic limit,  such as current reversals and  non-monotonic dependence of the diffusion coefficient on the temperature.

This work is organized as follows. In section~\ref{sec:model} we state the equation of motion of the overdamped Brownian particle and specify the model for the out-of-equilibrium substrate. In section~\ref{sec:deterministic} we explore the behavior of the system at the deterministic limit and we show that no mechanical rectification occurs in the deterministic limit.  In section~\ref{sec:adiabatic} we study the particle current of the model in the adiabatic limit, i.e., in the case in which the period of the external forcing is large compared to any typical time of the system. We give an analytical formula for the particle current based on recent exact results for disordered systems. In section~\ref{sec:numerical} we perform numerical simulations for the system in order to explore the transport characteristics beyond the adiabatic limit. Finally in section~\ref{sec:conclusions} we give a summary of our results and the conclusions of our work.

%$$$$$$$$$$$$$$$$$$$$$$$$$$$$$$$$$$$$$$$$$$$$$$$$$$$$$$$$$$$$$$
%$$$$$$$$$$$$$$$$$$$$$$$$$$$$$$$$$$$$$$$$$$$$$$$$$$$$$$$$$$$$$$
\section{Model}
\label{sec:model}
%$$$$$$$$$$$$$$$$$$$$$$$$$$$$$$$$$$$$$$$$$$$$$$$$$$$$$$$$$$$$$$
%$$$$$$$$$$$$$$$$$$$$$$$$$$$$$$$$$$$$$$$$$$$$$$$$$$$$$$$$$$$$$$

Let us consider an ensemble of Brownian particles moving on a given substrate. We assume that every Brownian particle only interacts with the substrate and that this interaction results in a potential $V(x)$. Thus, the  equation of motion that rules the dynamics of this particle is the following stochastic differential equation,
\begin{equation}
\label{eq:langevin}
\gamma dX_t = \left( f(X_t)+ F(t)\right) dt + \varrho_0 dW_t.
\end{equation}
In the above equation, $X_t$ represents the particle position at time $t$ and $f(x) = -V^\prime (x)$ corresponds to the force resulting from the interaction of the particle with the substrate. Additionally,  $W_t$ is a standard Wiener process modeling the thermal fluctuation and $F(t)$ is a time-dependent periodic external force.  The constants $\varrho_0^2$, and $\gamma $ are the noise intensity and the friction coefficient respectively. According to the fluctuation-dissipation theorem $\varrho_0^2 = 2 \gamma \beta^{-1}$, where $\beta$, as usual, stands for the inverse of the absolute temperature $\theta$ times the Boltzmann constant, $\beta = 1/k_B \theta$, .

Now let us describe the model for the substrate which was introduced in Refs.~\cite{rsg2013normal,rsg2014effective}. First we assume that the (one-dimensional) substrate is divided into ``cells'' of size $L$. Every cell can be thought of as a monomer which interacts with the particle via some interacting potential. We assume that the monomers comprising the substrate (called hereafter ``polymer'') can be of different types, and that all the possible types of monomers is finite, just as it occurs in a DNA molecule. Let $\mathcal{A}$ be the set of possible monomer types and let $\mathcal{A}^{\mathbb{Z}}$ be the set of all the possible polymers made up from monomers in $\mathcal{A}$. Then, a (random) polymer is represented by a symbolic sequence $\mathbf{a} \in \mathcal{A}^{\mathbb{Z}}$ which is of the form $\mathbf{a} =  (\dots,a_{-1},a_0,a_1,a_2,\dots)$ where $a_j$ is an element in $\mathcal{A}$ for all $j\in \mathbb{Z}$.  

\begin{figure}[t]
\begin{center}
\scalebox{0.35}{\includegraphics{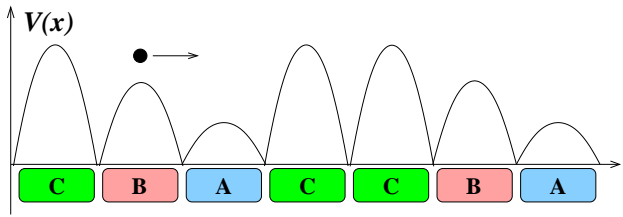}}
\end{center}
     \caption{
     Schematic representation of the model. The substrate, on which the particle moves, consists of a sequence of tracks of fixed length. On every track (called monomer) we define a potential profile which is interpreted as the potential interaction of the particle with the monomer. The letters $A$, $B$ and $C$ are displayed to indicate the monomer type on every track. The monomer type defines a unique potential profile and the disorder in the potential comes from the disorder of the monomer types along the substrate. The order of the monomers gives rise to a disordered medium that can be an out-of-equilibrium structure if the monomers are assembled according to an out of equilibrium process.
     }
\label{fig:particle-polymer}
\end{figure}

For the sake of simplicity, we assume that the particle interacts only with the closest monomer, i.e., the monomer at which the particle is located. Thus, the potential profile only depends on the monomer type on which the particle is located. See Figure~\ref{fig:particle-polymer} for a schematic representation of the model. Let us call $\psi(y,a)$ the interaction potential induced when the particle is located at the position $y\in[0,L)$ along the monomer of type $a\in\mathcal{A}$. Thus, if we write $x$ as $y+nL$ for some $n\in \mathbb{Z}$, the potential $V(x)$ that the particle ``feels'', can be explicitly written as
\begin{equation}
V(x) = \psi(y,a_n),
\end{equation} 
where $n$ labels the unit cell at which the particle is located and $y$ represents the relative position of the particle on the monomer. The symbol $a_n$ stands for type of the $n$th monomer on the chain. Analogously, we will denote by $\phi(y,a_n)$ the force field induced by $\psi(y,a_n)$, i.e., $f(x):= -V^\prime(x) = \phi(y,a_n)$, with $x = y + nL$, or equivalently,   $\phi(y,a_n) := -\psi^\prime(y,a_n)$, where the ``prime'' stands for the derivative with respect to the variable $y$. 

Now, let us state the model for the disordered substrate. In order to meet the condition that the polymer has an out-of-equilibrium structure we assume that the polymer is randomly produced by a Markov chain attaining a non-equilibrium stationary state (NESS). Thus a polymer $\mathbf{a} = (\dots,a_{-1},a_0, a_1,a_2,\dots)$, with $a_j \in \mathcal{A}$, is interpreted as a realization of a sequence of random variables $\{M_j \, :\, j\in\mathbb{Z}\}$ with joint probabilities $\mathbb{P}(M_0=a_0,M_1=a_1,\dots M_n=a_n) =: \mathbb{P}(a_0,a_1,\dots a_n) $, defined through a Markov matrix $\mathbf{Q}$ and its corresponding invariant probability row vector $ \boldsymbol{\pi} $ as follows,
\begin{equation}
\label{eq:markov-chain}
\mathbb{P}(a_0,a_1,\dots a_n)  := \pi(a_0) Q(a_0,a_1)  \dots Q(a_{n-1},a_n),
\end{equation}
for all $n\in \mathbb{Z}$. Notice that within the language of Markov chains, the set of possible monomer types $\mathcal{A}$ is called the state space, and the spatial variable $n$, indexing the monomers along the polymer, corresponds to the time variable of the stochastic process. The assumption that the Markov chain attains a NESS  implies that, if we draw a random polymer then a finite sequence and its reversal will not occur, in general, with the same probability along the polymer.  This property is what we call the \emph{spatial irreversibility } of the polymer.
This property can be explicitly written as follows. Given a finite sequence $\mathbf{a} = a_1a_2\dots a_n$, with $a_j \in \mathcal{A}$, we have that the probability that the reversed trajectory, $\bar{\mathbf{a}} = a_na_{n-1}\dots a_1$, occurs in the process is not the same as the probability that $\mathbf{a}$ occurs, i.e., we have in general that
\begin{equation}
\mathbb{P}(a_1,a_2,\dots,a_n) \not=\mathbb{P}(a_n,a_{n-1},\dots,a_1).
\end{equation}
The entropy production $e_p$ for the process $\{M_j \, :\, j\in\mathbb{Z}\} $ actually measures, in some way, the degree of time-irreversibility the process. In our context, the entropy production is a measure of the spatial irreversibility of a random polymer. This quantity can be defined as~\cite{maes1999fluctuation,maes2000definition},
\begin{equation}
\label{eq:ep_def}
e_p:= \lim_{n\to \infty} \ln\left( \frac{\mathbb{P}(a_1,a_2,\dots,a_n)  }{\mathbb{P}(a_n,a_{n-1},\dots,a_1)  } \right).
\end{equation} 
Particularly, it is known that for Markov chains the entropy production can be obtained directly by means of the corresponding Markov matrix~\cite{jiang2004mathematical}. If the Markov matrix is represented by $ \mathbf{Q}$ and its corresponding stationary probability vector is denoted by $ \boldsymbol{\pi} $, then, 
\begin{equation}
\label{eq:ep_markov}
e_p = \frac{1}{2}\sum_{a,b\in\mathcal{A}} \left( \pi (a) Q(a,b) - \pi (b)Q(b,a) \right)\ln \left(\frac{ \pi(a) Q(a,b)}{\pi(b) Q(b,a)} \right).
\end{equation}
Next we chose a specific model of Markov chain having a non-equilibrium stationary state, in which the entropy production can be tuned by varying a single parameter. For this purpose we assume that the state space  is  $\mathcal{A}:=\{0,1,2\}$, i.e., there are only three monomer types labeled by the symbols $0$, $1$ and $2$. The Markov chain is defined through the one-parameter Markov matrix $\mathbf{Q}$, given by,
\begin{equation}
\label{eq:Q}
\mathbf{Q} = \left(
  \begin{array}{ccc}
   0 & p & 1-p \\
   1-p & 0 & p \\
   p & 1-p & 0
  \end{array} \right).
\end{equation}
It is easy to check that the matrix $\mathbf{Q}$ is doubly stochastic and has a unique invariant probability vector $\boldsymbol{\pi} = \boldsymbol{\pi} \mathbf{Q} $, given by $\boldsymbol{\pi} = (\frac{1}{3},\frac{1}{3},\frac{1}{3})$. This model attains a stationary state which is of equilibrium for the parameter value $p=1/2$. For $p\not=1/2$ the stationary state is of non-equilibrium and its entropy production is given by~\cite{jiang2004mathematical}
\begin{equation}
\label{eq:ep_markov_p}
e_p = (2p-1)\ln\left( \frac{p}{1-p}\right). 
\end{equation}
It is important to note that, independently of the parameter value $p$, the probability vector $\boldsymbol{\pi}$ is always the same. In other words, monomer types in a typical realization of the random polymer are equally distributed independently of the value of the entropy production. In the following we will study the dynamics of an ensemble of Brownian particles moving on the above-described out-of-equilibrium structures.

%%==================== FIGURE =========================
%%
%%==================== FIGURE =========================
%

%$$$$$$$$$$$$$$$$$$$$$$$$$$$$$$$$$$$$$$$$$$$$$$$$$$$$$$$$$$$$$$
%$$$$$$$$$$$$$$$$$$$$$$$$$$$$$$$$$$$$$$$$$$$$$$$$$$$$$$$$$$$$$$
\section{Deterministic limit}
\label{sec:deterministic}
%$$$$$$$$$$$$$$$$$$$$$$$$$$$$$$$$$$$$$$$$$$$$$$$$$$$$$$$$$$$$$$
%$$$$$$$$$$$$$$$$$$$$$$$$$$$$$$$$$$$$$$$$$$$$$$$$$$$$$$$$$$$$$$

In this section we consider the dynamics of our model in absence of noise. In this case, the dynamics of a particle on a random potential is governed by the equation
\begin{equation}
\gamma \frac{dx}{dt} =  f(x) + F(t).
\label{eq:deterministic}
\end{equation}
Here  $f(x)$ is minus the gradient of the potential $V(x) := \psi(a_n,y)$ (where $x = nL+y$) defined above that depends on a realization of the random polymer $\mathbf{a} = (\dots, a_{-1},a_0,a_1,a_2,\dots)$, with $a_j \in \{0,1,2\}$. The function $F(t)$ is a time-periodic external force with period $T$. Throughout this work we will use a simple form for $F(t)$, 
\begin{equation}
F(t) =  \left\{ \begin{array} 
            {r@{\quad \mbox{ if } \quad}l} 
   F_0   &   0\leq t~\mbox{mod}~[T] < T/2   \\ 
   -F_0  &   T/2 \leq t~\mbox{mod}~[T]  < T.          \\ 
             \end{array} \right. 
\label{eq:Ft}
\end{equation}
This choice for $F(t)$ allows us to analyze the trajectories described by Eq.~\eqref{eq:deterministic} by means of the technique developed in Refs.~\cite{rsg2006deterministic,rsg2008occurrence}. Actually, our goal is to prove that, independently of the initial condition and independently of the realization of the polymer (random potential), the solution to Eq.~\eqref{eq:deterministic}, $x(t)$, does not diverge in time. The main hypothesis we use for the potential profile $\psi(y,a_n)$ is that it is symmetric in the sense ratchet systems~\cite{reimann2002brownian}. This condition establishes that the potential $\psi(y;a)$ is symmetric on $[0,L]\subset\mathbb{R}$ if the force field $\phi (y,a_n) := -\psi^\prime(y,a) $  satisfy that,
\begin{eqnarray}
\phi (y,a_n) = -\phi (L-y,a_n).
\end{eqnarray}

As we said above, we use the technique developed in~\cite{rsg2006deterministic,rsg2008occurrence} to prove the absence of rectification phenomenon in this system. However, such a technique does not apply directly to our case because the potential we use is not periodic. Actually in~\cite{rsg2006deterministic} it was proved that the dynamics of an overdamped particle in a periodic potential and under the influence of a time-dependent periodic forcing,  can be described by sampling periodically the position. The result is a discrete-time trajectory that is ruled by a circle map (specifically, a \emph{lift} of a circle homeomorphism). In our case, although we can still perform a sampling at regular time-intervals of the continuous trajectory to generate a discrete one, we cannot obtain a single mapping to reproduce the particle motion.  This is of course, a consequence of the disorder of the potential. 

Let $x(t)$ be a solution of Eq.~\eqref{eq:deterministic} with initial condition $x(0)=x_0$. We define the sequence  $\{x_n \in \mathbb{R} \,: \, n\in \mathbb{N}\}$ such that $ x_n := x(nT/2)$. We can see that $x_n$ corresponds to the particle position at the beginning of every half of the period of $F(t)$. Next we define a sequence $\{y_n \in [0,1)  \,: \, n\in \mathbb{N}\}$ of \emph{reduced positions}  as follows, $y_n = x_n \mod [L]$. Recall that $L$ is the length of the ``unit cell'', i.e., the length of every monomer.  We can say alternatively that the position $x_n$ of the particle can be written as $ x_n = y_n + m_n L$ where $y_n$ (the \emph{reduced position}) corresponds to the particle position relative to the monomer at which it is located. The integer $m_n$ is the monomer where the particle is found at $t = nT/2$. With these definitions we can say that the particle motion can split in two parts, $i$) a sequence of integers labeling the monomers that the particle has visited $\{ m_n \in \mathbb{Z} \, : \, n\in \mathbb{N}\}$,  and $ii$) a sequence of numbers indicating the relative particle position on the visited monomer $\{y_n \in [0,1)  \,: \, n\in \mathbb{N}\}$. The former can be thought of as a coarse-grained description of the trajectory of $x(t)$, giving information on the monomers that the particle has visited, while the latter is interpreted as the relative position of the particle with respect to the monomer (or unit cell). The proof of the absence of the rectification phenomenon consist of two parts. Firstly, it is necessary to show that there exists an integer  $n^* \in \mathbb{N}$ such that $m_{n^*} = m_{n^* + 2}$. This means that if a $t= n^* T/2$ the particle is located at the $m_{n^*}$th track (at the $m_{n^*}$th  monomer) then at $ t =  n^* T/2 + T$ (after one period) the particle will return to the same unit cell. Next, we can prove that, once the particle departs and returns to the same monomer after one period, the particle remains in this dynamics, i.e., the particle gets trapped by a kind of ``coarse-grained periodic orbit''.  The last statements actually implies that the mapping $R$ ruling the behavior of the discrete trajectory $\{x_n : n\in \mathbb{N}\}$ has at least one fixed point, according to a theorem about non-decreasing maps on an interval~\cite{katok1997introduction}. The existence of such a fixed point for the map $R$ actually means that the full trajectory of the particle $x(t)$, eventually reaches a periodic orbit. All the above-mentioned proofs are given in Supplemental Material (SM)~\footnote{See Supplemental Material at [URL will be inserted by publisher] for the proof of absence of the rectification phenomenon in absence of noise.}.

%%==================== FIGURE =========================
%%
\begin{figure}[t]
\begin{center}
\scalebox{0.4}{\includegraphics{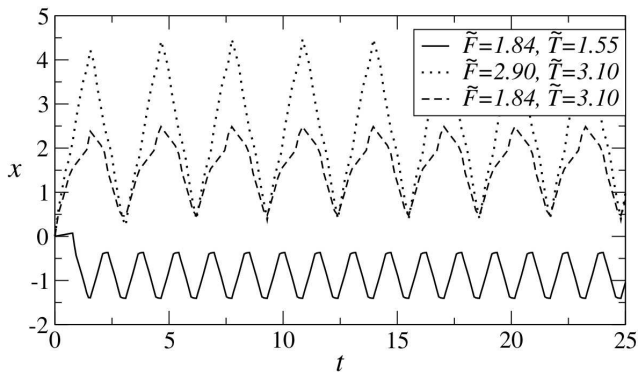}}
\end{center}
     \caption{
     Realizations of the particle position $x(t)$ in the deterministic limit for several values of the dimensionless parameters $\tilde F$ and $\tilde T$ (see Section~\ref{sec:adiabatic} for the definition of dimensionless parameters). The parameter $p$ for the Markov chain was chosen as $p=0.99$, which is very close to the totally irreversible case $p=1$. We should notice that, despite the high statistical asymmetry in the random potential, we have no particle current arising in this case. Here ``high statistical asymmetry'' means that the random potentials were produced by a process with an entropy production near its maximal value.   
          }
\label{fig:orbits-zero-temperature}
\end{figure}
%%==================== FIGURE =========================
%

In Figure~\ref{fig:orbits-zero-temperature} we display some trajectories of the system at zero noise strength. We observe that these trajectories always reach a periodic orbit, showing that the rectification phenomenon does not occur. This is in agreement with the above claim, which states that a particle eventually reaches a periodic orbit independently of the disorder and the realization of the disordered potential.

Thus we have proven that dynamics of the model in absence of noise exhibits a zero particle current. This means that the statistical  asymmetry of the substrate is not a sufficient condition for the system to exhibit the rectification phenomenon at the deterministic limit, as it occurs in ratchet systems~\cite{rsg2006deterministic,rsg2008occurrence,reimann2002brownian, bartussek1994periodically,Mateos2000chaotic,Arzola2011experimental}. Indeed, the fact that the system is unable to rectify motion of particles in absence of noise is due to the symmetry of the potential profile on every unit cell which is the main hypothesis done to achieve the proofs.

%$$$$$$$$$$$$$$$$$$$$$$$$$$$$$$$$$$$$$$$$$$$$$$$$$$$$$$$$$$$$$$
%$$$$$$$$$$$$$$$$$$$$$$$$$$$$$$$$$$$$$$$$$$$$$$$$$$$$$$$$$$$$$$
\section{Particle current in the adiabatic limit}
\label{sec:adiabatic}
%$$$$$$$$$$$$$$$$$$$$$$$$$$$$$$$$$$$$$$$$$$$$$$$$$$$$$$$$$$$$$$
%$$$$$$$$$$$$$$$$$$$$$$$$$$$$$$$$$$$$$$$$$$$$$$$$$$$$$$$$$$$$$$

As mentioned in Section~\ref{sec:model}, we are considering the dynamics of an overdamped particle on a disordered potential, $V(x)  = \psi (y,a_n)$, subjected to a time-dependent periodic driving force  $F(t)$, which is ruled by the stochastic differential equation,
\begin{equation}
\label{eq:langevin-again}
\gamma dX_t = \left( f(X_t)+ F(t)\right) dt + \varrho_0 dW_t.
\end{equation}
Here we assume that the potential profile $\psi(y,a_n)$ at the $n$th unit cell is a piece-wise linear function of the form
\begin{equation}
\psi(y,a_n)  =  \left\{ \begin{array} 
            {r@{\quad \mbox{ if } \quad}l} 
  \alpha (a_n) y   &  0\leq  y<L/2    \\ 
   \alpha (a_n)(L-y )  &  L/2 \leq y < L,      \\ 
             \end{array} \right.
\label{eq:potential-model}
\end{equation}
where $\alpha : \mathcal{A} \to \mathbb{R}$ is a function giving the slopes of the potential profile depending of the monomer type at which the particle is located.  This form of the potential profile complies with the symmetry criterion given in Sec.~\ref{sec:deterministic} and will be useful to obtain a explicit expression for the particle current in the adiabatic limit.
As stated above, we will assume that the time-dependent periodic forcing is a square-wave function given by,
\begin{equation}
F(t) =  \left\{ \begin{array} 
            {r@{\quad \mbox{ if } \quad}l} 
   F_0   &   0\leq t~\mbox{mod}~[T] < T/2   \\ 
   -F_0  &   T/2 \leq t~\mbox{mod}~[T]  < T.          \\ 
             \end{array} \right. 
\label{eq:Ft}
\end{equation}
We should notice that $F(t)$ is a periodic function with period $T$.  During every half  of the period, the external driving force remains constant and the system is described by means of a time-independent stochastic differential equation. During the first half of the period, the external forcing is $F_0$ and the dynamic equation is,
\begin{equation}
\label{eq:adiabatic+}
\gamma dX_t = \left( f(X_t)+ F_0\right) dt + \varrho_0 dW_t.
\end{equation}
During  the second half of the period, the external driving force turns into $-F_0$ and the system is described by the equation
\begin{equation}
\label{eq:adiabatic-}
\gamma dX_t = \left( f(X_t)- F_0\right) dt + \varrho_0 dW_t.
\end{equation}
It is clear then, that the set of equations given by expressions~\eqref{eq:adiabatic+} and~\eqref{eq:adiabatic-} are completely equivalent to Eq~\eqref{eq:langevin-again}.

We are interesting in studying the behavior of particle current $J_{\mathrm{eff}}$  and the effective diffusion coefficient $D_{\mathrm{eff}}$ for this system. These quantities are defined as
\begin{eqnarray}
\label{eq:def_Jeff}
J_{\mathrm{eff} } &:=& \lim_{t\to \infty} \frac{\langle \langle  X_t\rangle\rangle  }{ t},
\\
D_{\mathrm{eff} } &:=& \lim_{t\to \infty} \frac{\langle \langle  X_t^2\rangle\rangle - \langle \langle  X_t\rangle\rangle }{2 t},
\end{eqnarray}
where the symbol $\langle \langle \cdot \rangle \rangle$ denotes a double average, the first one taken with respect to noise (maintaining fixed the disorder of the polymer), and the second one taken over an ensemble of disordered polymers~\cite{rsg2014effective}. 

It is easy to see that $J_{\mathrm{eff} }$ can be written as,
\begin{eqnarray}
J_{\mathrm{eff} } &=& \lim_{n \to \infty} \frac{\langle \langle  X_{nT/2}\rangle\rangle  }{ nT} ,
\nonumber
\\
&=& \lim_{n \to \infty} \frac{ 1 }{ nT} \sum_{j=1}^n \left( \langle \langle  X_{jT/2}\rangle\rangle - \langle \langle  X_{(j-1)T/2}\rangle\rangle \right),
\label{eq:Jeff-1}
\end{eqnarray}
or, equivalently,
\begin{eqnarray}
J_{\mathrm{eff} } 
&=& \frac{1}{2} \bigg( \lim_{n \to \infty} \frac{ 1}{ n} \sum_{j \ \mathrm{ odd}}^n \frac{ \langle \langle  X_{jT/2}\rangle\rangle - \langle \langle  X_{(j-1)T/2}\rangle\rangle}{T/2} 
\nonumber
\\
&+& 
\lim_{n \to \infty} \frac{ 1 }{ n} \sum_{j \ \mathrm{ even}}^n 
\frac{ \langle \langle  X_{jT/2}\rangle\rangle - \langle \langle  X_{(j-1)T/2}\rangle\rangle }{T/2} \bigg).
\label{eq:Jeff-2}
\end{eqnarray}
Next, if we assume that the period $T$ is large with respect to any relaxation time of the system, assuming constant driving force on every half of the period, it is clear that the particle current given in Eq.~\eqref{eq:Jeff-2} can be written as
\begin{equation}
\label{eq:def_Jeff-adiabatic}
J_{\mathrm{eff} } =  \frac{1}{2} \left( J_\mathrm{s}(+F_0) + J_\mathrm{s}(-F_0) \right)
\end{equation}
where $J_\mathrm{s} (F_0)$ is defined as the particle current in the stationary state of the system subjected to a constant driving force $F_0$. The stochastic differential equation by means of which we can obtain $J_\mathrm{s} $ is written as follows,
\begin{equation}
\label{eq:adiabatic}
\gamma dX_t = \left( f(X_t)+ F_0\right) dt + \varrho_0 dW_t.
\end{equation}
The formula~\eqref{eq:def_Jeff-adiabatic} for the particle current for the time-dependent system is commonly known as the \emph{adiabatic approximation}.

In Ref.~\cite{rsg2014effective} the author has obtained the exact expression for the particle current (and the diffusion coefficient) of overdamped particles moving on disordered potentials subjected to a constant driving force governed by Eq.~\eqref{eq:adiabatic}. Using such a formula we obtain that 
\begin{eqnarray}
J_\mathrm{s} (F) &=& \frac{L}{\langle T_{1}  (\mathbf{a};F_0)  \rangle_{\mathrm{p}}}.
\label{eq:Js}
\end{eqnarray}
Here, the notation  $\langle \cdot \rangle_{\mathrm{p}}  $ means average with respect to the polymer ensemble and  $T_1(\mathbf{a};F_0)$ stands for the mean first passage time (MFPT) from $x=0$ to $x=L$ for the process defined by the stochastic differential equation~\eqref{eq:adiabatic}. It is clear that $T_1$ depends on the disorder (which is represented by the symbolic sequence $\mathbf{a}$) and the external driving force $F_0$.  Actually there are standard techniques to calculate the MFPT in terms of quadratures~\cite{risken1996fokker,hanggi1990reaction}. In Ref.~\cite{rsg2014effective} it was shown that for the case of disordered potentials the MFPT can be written as, 
\begin{eqnarray}
 T_1(\mathbf{a};F_0) &=& \gamma \beta \sum_{m=1}^\infty e^{-m\beta F_0 L} q_{+}(a_0)q_{-}(a_m) + \gamma \beta I_0(a_0),
 \nonumber
 \\ 
 &&
\label{eq:T1}
\end{eqnarray}
where the functions $q_+$ and $q_-$ and $I_0$ are defined as,
\begin{eqnarray}
\label{eq:def-qpm}
q_\pm (a) &:=& \int_0^L e^{ \pm \beta [ \psi(x,a) - xF_0 ] } dx ,
\\
I_0 (a)  &:=&  \int_0^L \int_0^x e^{ - \beta [ \psi(y,{a}) -  \psi(x,{a}) + (x-y)F_0]}\, dy \, dx. \qquad
\end{eqnarray}

Recall that the potential profile $\psi (y,a)$ has been chosen as a piece-wise linear function as we can see in Eq.~\eqref{eq:potential-model}. This choice is particularly convenient to obtain an explicit expression for all the quantities involved in the MFPT given by Eq.~\eqref{eq:T1}. Thus, using the above mentioned model for the potential profile, it is not hard to see that the functions $q_\pm$ and $I_0$ can be written as,
\begin{eqnarray}
q_\pm (a) &=&  \frac{L}{2}\mathcal{E}\left(\pm \beta (\alpha(a)-F_0)L/2   \right) 
\nonumber 
\\
&+& \frac{L}{2}e^{-\beta F_0 L }  \mathcal{E}\left(\pm \frac{\beta L}{2} (\alpha(a)+F_0)   \right),
\\
I_0(a) &=& \frac{L}{2 \beta} \frac{2 F_0}{F_0^2 - \alpha^2(a)}  - \frac{L}{2 \beta} \frac{\mathcal{E}\left( \beta (F_0 - \alpha(a))L/2 \right)}{F_0 - \alpha(a)} 
\nonumber
\\
&-& \frac{L}{2 \beta} \bigg( \frac{e^{\beta \alpha(a) L}}{F_0+\alpha(a)}\bigg) \mathcal{E}\left( -\frac{\beta L}{2} (F_0 + \alpha(a)) \right)
\nonumber
\\
&+&  \frac{ L^2}{4} e^{\beta (F_0-\alpha(a))L}\mathcal{E}\left( \frac{\beta L}{2} (F_0 - \alpha(a)) \right)  
\nonumber
\\
&\times& \mathcal{E}\left(- \frac{\beta L}{2} (F_0 + \alpha(a)) \right).
\end{eqnarray}
In the above expressions we made use of the function $\mathcal{E} : \mathbb{R} \to \mathbb{R} $ which is defined as
\[
\mathcal{E} (x)  := \frac{e^{x}-1}{x}.
\]

Once we have the explicit expressions for the functions involved in Eq.~\eqref{eq:T1} for the MFPT, we need to establish the manner in which we perform the average with respect to the polymer ensemble, 
\begin{eqnarray}
 \langle T_1(\mathbf{a};F_0) \rangle_{\mathrm{p}} &=& \gamma \beta \sum_{m=1}^\infty e^{-m\beta F_0 L} \langle q_{+}(a_0)q_{-}(a_m) \rangle_{\mathrm{p}} 
 \nonumber
 \\ 
 &+&\gamma \beta \langle I_0(a_0)  \rangle_{\mathrm{p}}.
\label{eq:T1_avr}
\end{eqnarray}
We should observe that it is necessary to take the average of two functions, one depending only on one monomer of $\mathbf{a}$, namely $I_0(a_0)$, and another one depending on two monomers, the function $q_{+}(a_0)q_{-}(a_m)$. To compute such averages we need the corresponding marginal distributions, i.e., we need to know explicit expressions for the probabilities $\mathbb{P}(a_0)$ and $\mathbb{P}(a_0,a_m) := \mathbb{P}(M_0 = a_0; M_m = a_m)$. 

As mentioned before, we assume that the sequence $\mathbf{a} = (\dots a_{-1}, a_0, a_1,a_2,\dots)$ is built up by means of a stationary Markov chain with Markov matrix $\mathbf{Q}$ given in Eq.~\eqref{eq:Q}. It is known that the one-dimensional marginal distribution $\mathbb{P}(a_0)$  for a Markov chain corresponds to the invariant probability vector $\boldsymbol{\pi}$, i.e., $\mathbb{P}(a_0) = \pi(a_0)$. On the other hand, the two-dimensional marginal distribution, $\mathbb{P}(a_0,a_m)$,  is obtained by means of the Markov matrix $\mathbf{Q}$ as follows,
\begin{equation}
\mathbb{P}(a_0,a_m) = \sum_{a_1}  \dots \sum_{a_{m-1}} \pi(a_0)Q(a_0,a_1) \dots Q(a_{m-1},a_m).
\end{equation}
where the summation runs over all possible ``states'' of the Markov chain, i.e., over all possible monomer types. With these expressions it is possible to calculate the following averages,
\begin{eqnarray}
\langle I_0(a_0)  \rangle_{\mathrm{p}} &=& \sum_{a_0} I_0(a_0) \pi (a_0)
\\
\langle q_{+}(a_0)q_{-}(a_m) \rangle_{\mathrm{p}} &=& \sum_{a_0} \sum_{a_m} q_{+}(a_0)q_{-}(a_m)  \mathbb{P}(a_0,a_m).\qquad
\end{eqnarray}
Up to now, all the above quantities have been written explicitly allowing us to numerically evaluate the particle current as a function of the parameters. 

Now we define dimensionless quantities in order to perform numerical experiments showing the validity of the theoretical results.  For this purpose let us start defining a ``critical driving force'' as the average of the slopes involved in the random potential, i.e., we define $\overline{\alpha} $ as,
\begin{equation}
\overline{\alpha} := \sum_{a=0}^2 \alpha(a) \pi (a).
\end{equation}
We also define a ``relaxation time''  $t_\mathrm{r}$ as follows,
\begin{equation}
t_\mathrm{r} := \frac{\gamma L }{\overline{\alpha} }  
\end{equation} 
With these quantities now we define dimensionless parameters as follows. First of all we define the dimensionless forcing amplitude, $\tilde{F}_0$, as $\tilde{F}_0 := F_0/\overline{\alpha}$. We also define the dimensionless period, $\tilde T $, of the time-dependent driving force as
\begin{equation}
\tilde T := \frac{T}{t_{\mathrm{r}}} = \frac{T \,\overline{\alpha} }{ \gamma L}. 
\end{equation}
Now we define a dimensionless temperature $\tilde \theta$ as follows,
\begin{equation}
\tilde \theta := \frac{\beta^{-1}}{\overline{\alpha} L} = \frac{k_\mathrm{B} \theta}{\overline{\alpha} L},
\end{equation}
where $k_\mathrm{B}$ is the Boltzmann constant and $\theta$ is the absolute temperature. Finally, we define a dimensionless particle current and a dimensionless effective diffusion coefficient as, 
\begin{eqnarray}
\tilde{J}_{\mathrm{eff}} &:=& \frac{{J}_{\mathrm{eff}}}{ L/t_{\mathrm{r}}} = \frac{\gamma {J}_{\mathrm{eff}}}{ \overline{\alpha }},
\\
\tilde{D}_{\mathrm{eff}} &:=& \frac{{D}_{\mathrm{eff}}}{ L^2/t_{\mathrm{r}}} = \frac{\gamma {D}_{\mathrm{eff}}}{ L \, \overline{\alpha }}.
\end{eqnarray}

At this point we can numerically evaluate the formula for the particle current~\eqref{eq:def_Jeff-adiabatic} if we specify  the values of the parameters. For this purpose let us state the parameter values that will be used for the numerical evaluation of the particle current in the adiabatic limit as well as for the numerical simulations of the Langevin equation.  Throughout the rest of this work we will take $\gamma = 1$, $L=1$, and the slopes of the random potential as $\alpha(0) = 0.1$, $\alpha(1) =1.2$ and $\alpha(2) = 1.8$. 
%
%
%
%%==================== FIGURE =========================
%%
\begin{figure}[t]
\begin{center}
\scalebox{0.4}{\includegraphics{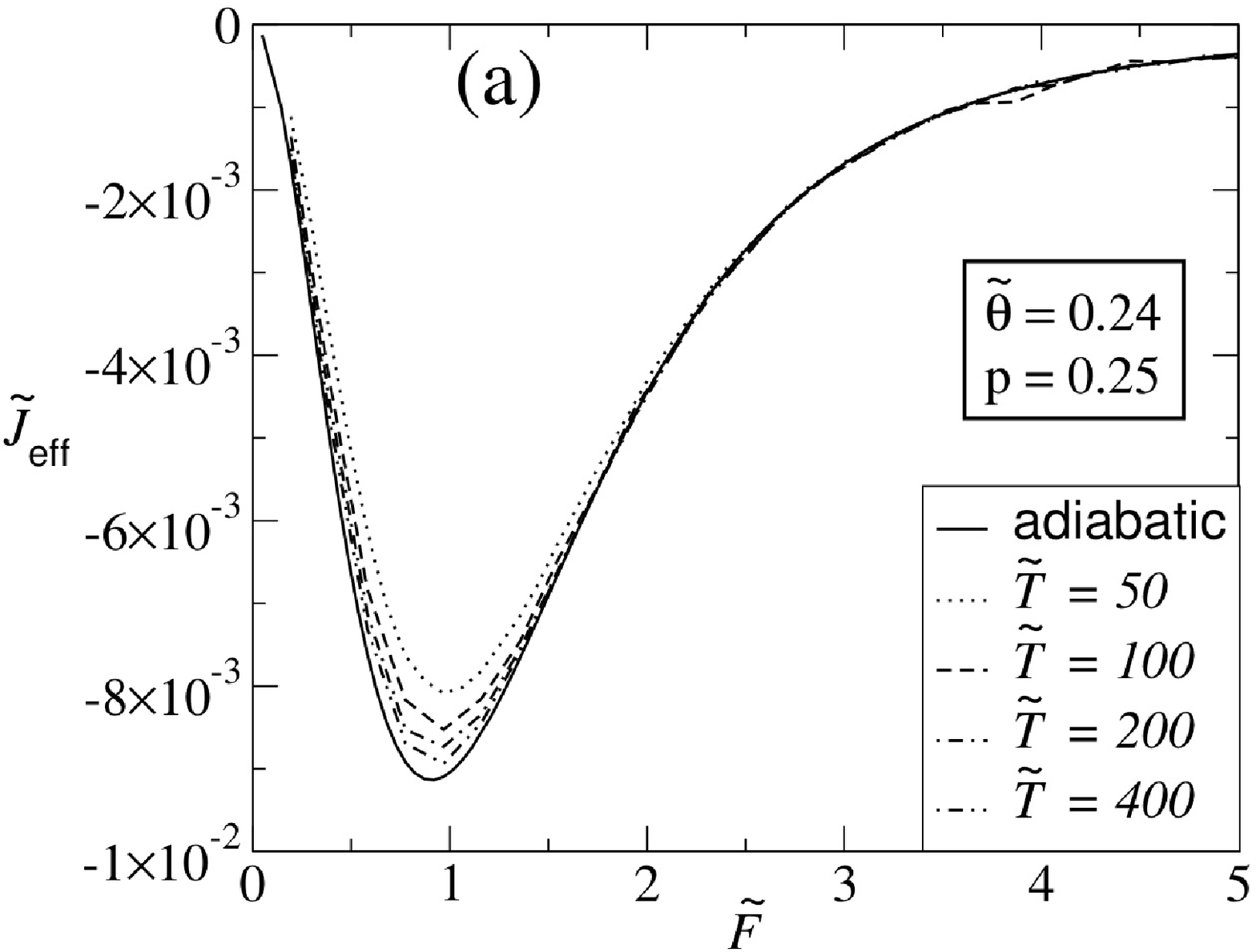}}
\scalebox{0.40}{\includegraphics{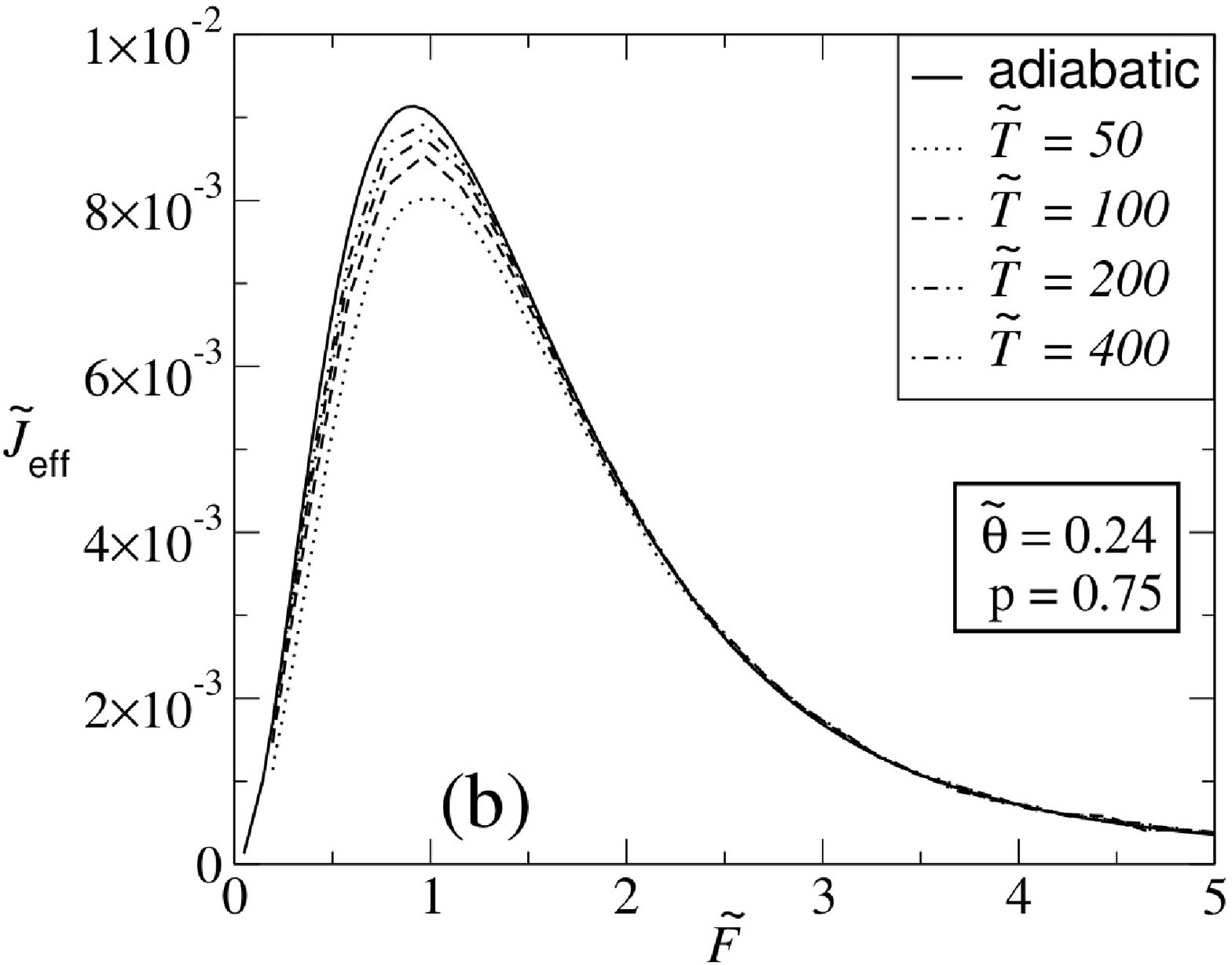}}
\end{center}
     \caption{
       (a) Particle current as a function of driving force. We display the particle current obtained from numerical simulations for several values of the dimensionless period $\tilde T$ and for the parameter values  $p=0.25$ and $\tilde \theta = 0.24$. We also plot the corresponding curve obtained from the theoretical prediction in the adiabatic limit ($\tilde T = \infty$) given by Eq.~\eqref{eq:def_Jeff-adiabatic}. We can appreciate that as the period increases the particle current approaches the adiabatic limit curve, in agreement with our theoretical predictions. (b) The same as (a)  for $p=0.75$. We should observe the difference in sign in the particle current.
     }
\label{fig:J_adiabatic}
\end{figure}
%%==================== FIGURE =========================
%
%
%
%
%

\begin{figure}[t]
\begin{center}
\scalebox{0.35}{\includegraphics{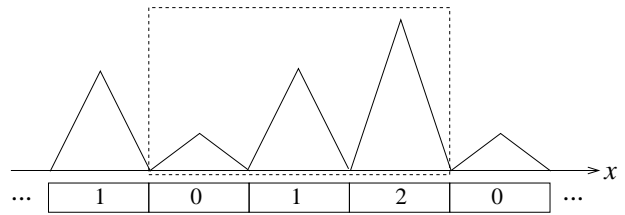}}
\end{center}
     \caption{
      Schematic representation of a realization of the random potential. The numbers 0,1 and 2 labeling the monomers are obtained by a realization of a three-states Markov chain. Each state translates into a potential profile having different heights. It is clear that the potential profile on a unit cell (a monomer) is  symmetric by definition. However, if we see the potential profile on three unit cells, the potential have a clear global asymmetry, as we can see in the consecutive monomers 0,1,2 shown in the figure.
           }
\label{fig:asymmetry}
\end{figure}

In figure~\ref{fig:J_adiabatic}a we plot the particle current in the adiabatic limit for $p=0.25$ and $ \tilde \theta = 0.24$ obtained by numerically evaluating Eq.~\eqref{eq:def_Jeff-adiabatic}. Figure~\ref{fig:J_adiabatic}a also shows the particle current for several values of the dimensionless period: $\tilde T = 50$ , $ \tilde T = 100$, $\tilde T = 200 $ and $\tilde T = 400$. We observe that the particle current curve approaches to the adiabatic limit curve as the period increases showing that the numerical simulations are consistent with our theoretical prediction. Analogously, in figure~\ref{fig:J_adiabatic}b we appreciate the convergence of the particle current curves, obtained by numerical simulations,  to the  theoretically predicted adiabatic limit curve for $p=0.75$. We should notice that the particle current is positive for $p=0.75$ and negative for $p=0.25$. This fact can be explained by comparing our model with a disordered ratchet model~\cite{marchesoni1997transport}. Although our system does not behaves as ratchet system in the limit of zero noise strength (because of the asymmetry and the vanishing particle current), we can still explain the sign of the particle current by thinking that the potential is ``asymmetric'' in a domain larger than the unit cell. This asymmetry, which in the context of non-equilibrium system is called irreversibility, can be observed in Fig.~\ref{fig:asymmetry} where we sketch a realization of the random potential. We should observe that the potential limited to a single unit cell is symmetric by definition. However, if extended the domain of observation, for instance, to three unit cells, we will see that the profile is asymmetric if we look at the appropriate unit cells. Recall that the potential profile associated to a realization of the polymer is made according to the function $\alpha $. This function was chosen in such a way that the highest profile is assigned to the monomer $2$ and the lower to the monomer $0$. Then, if a given realization prefers the three-monomer structures $012$, then the potential profile will look like an ``asymmetric effective potential''  as can be appreciated in Fig.~\ref{fig:asymmetry}. This ``effective potential'', in the context of rocket thermal ratchets, would rectify to the right~\cite{reimann2002brownian}. It is not difficult to see that the Markov chain we have chosen, the  the three-monomer structures $012$ will occur with higher probability than the  the three-monomer structures $210$ if $p>1/2$. The latter means that for $p>1/2$ the random potential will have a potential profile with a preferred asymmetry rectifying to the right, as it has been demonstrated in the case of disordered ratchets~\cite{marchesoni1997transport}. This is consistent with the fact that the particle current is positive for $p>1/2$ as shown in Fig.~\ref{fig:J_adiabatic}b. The same argument applies for the case $p<1/2$, which implies a negative particle current, as shown in Fig.~\ref{fig:J_adiabatic}a. Notice that, despite the irreversibility of the process give rise a kind of ``coarse grained asymmetry''  in the random potential, this asymmetry is not enough to induce a ratchet effect  at the deterministic limit as we have already proven.
%
%
%%==================== FIGURE =========================
%%
%%==================== FIGURE =========================
%
%
%

\begin{figure}[t]
\begin{center}
\scalebox{0.35}{\includegraphics{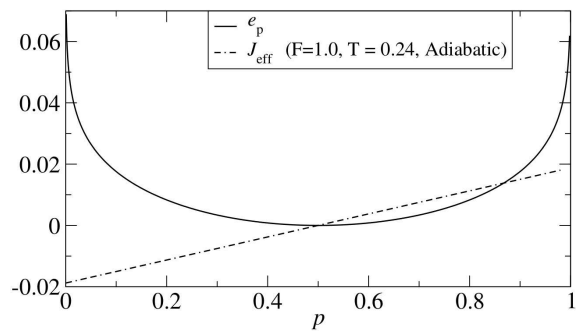}}
\end{center}
     \caption{
      Particle current and entropy production. We compare the behavior of $J_\mathrm{eff}$ (solid line) as a function of $p$ with the entropy production of the substrate (dot-dashed line). We observe that the minimum entropy production is zero and is attained at $p=1/2$. For such a parameter value we have that the structure is produced under equilibrium conditions and the particle current is zero. We also observe that, the larger entropy production, the larger particle current (in absolute value).
     }
\label{fig:J-vs-entropy}
\end{figure}

It is also important to observe how the particle current behaves as a function of the parameter $p$ of the Markov chain that generates the disordered substrate. We should recall that $p$  controls, in some way, the degree of irreversibility of the ``polymer''. Actually, for $p=1/2$ the substrate is generated under equilibrium conditions and the irreversibility grows as $p$ approaches the extreme values $p=0$ or  $p=1$. In Fig.~\ref{fig:J-vs-entropy} we show the behavior of $\tilde J_{\mathrm{eff}}$ as a function of $p$. To compare this behavior against the entropy production, we also plot $e_p$ as a function of $p$, but scaled by a constant factor of $ 10^{-3}$ just to compare it with the particle current. We see that the minimum occurs at $p=1/2$ for the particle current as well as for the entropy production.

%
%
%%==================== FIGURE =========================
%%
%%==================== FIGURE =========================
%
%
%

%$$$$$$$$$$$$$$$$$$$$$$$$$$$$$$$$$$$$$$$$$$$$$$$$$$$$$$$$$$$$$$
%$$$$$$$$$$$$$$$$$$$$$$$$$$$$$$$$$$$$$$$$$$$$$$$$$$$$$$$$$$$$$$
\section{Particle current and diffusion coefficient for fast tilting}
\label{sec:numerical}
%$$$$$$$$$$$$$$$$$$$$$$$$$$$$$$$$$$$$$$$$$$$$$$$$$$$$$$$$$$$$$$
%$$$$$$$$$$$$$$$$$$$$$$$$$$$$$$$$$$$$$$$$$$$$$$$$$$$$$$$$$$$$$$

\begin{figure}[t]
\begin{center}
\scalebox{0.44}{\includegraphics{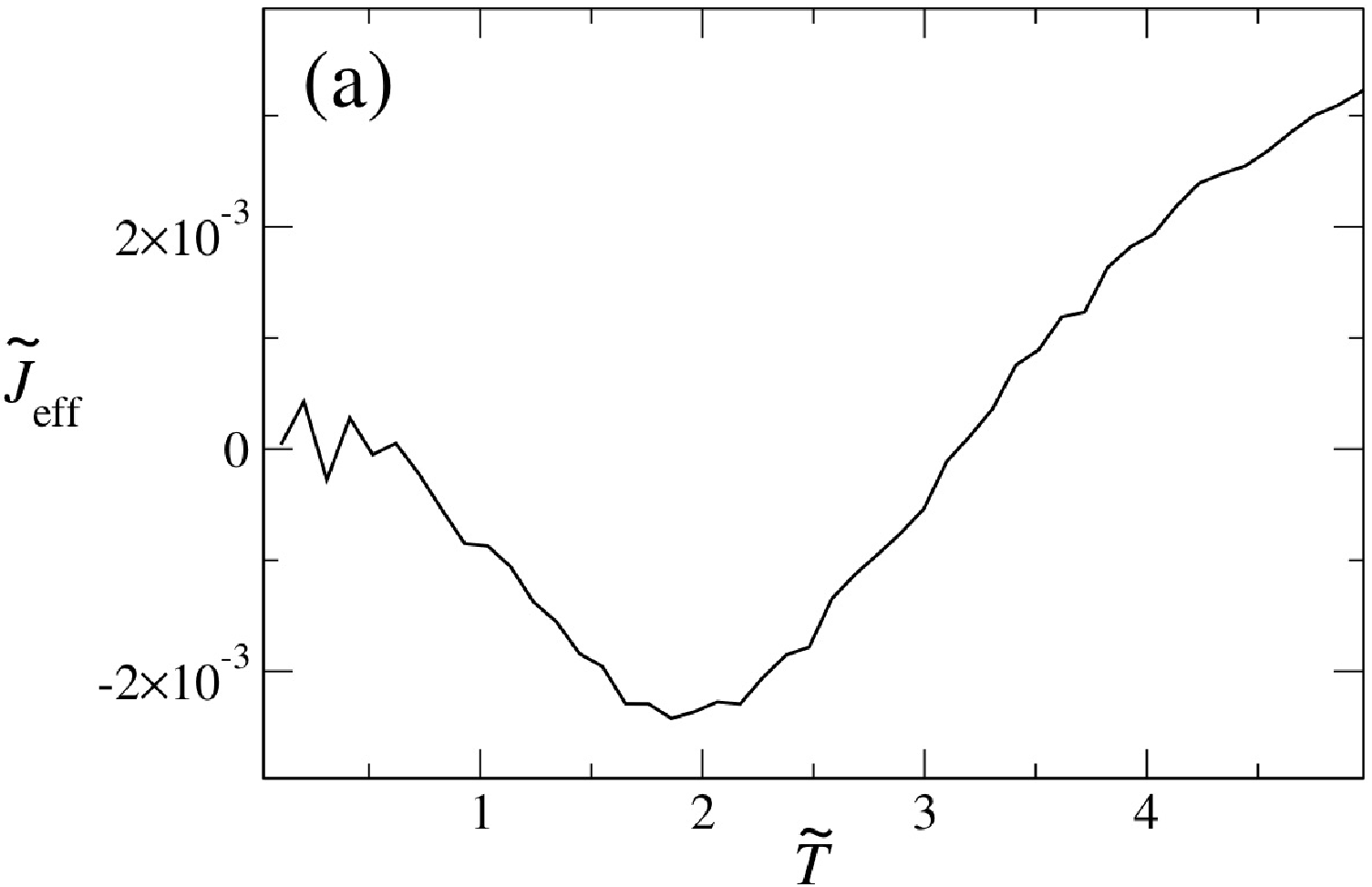}}
\scalebox{0.42}{\includegraphics{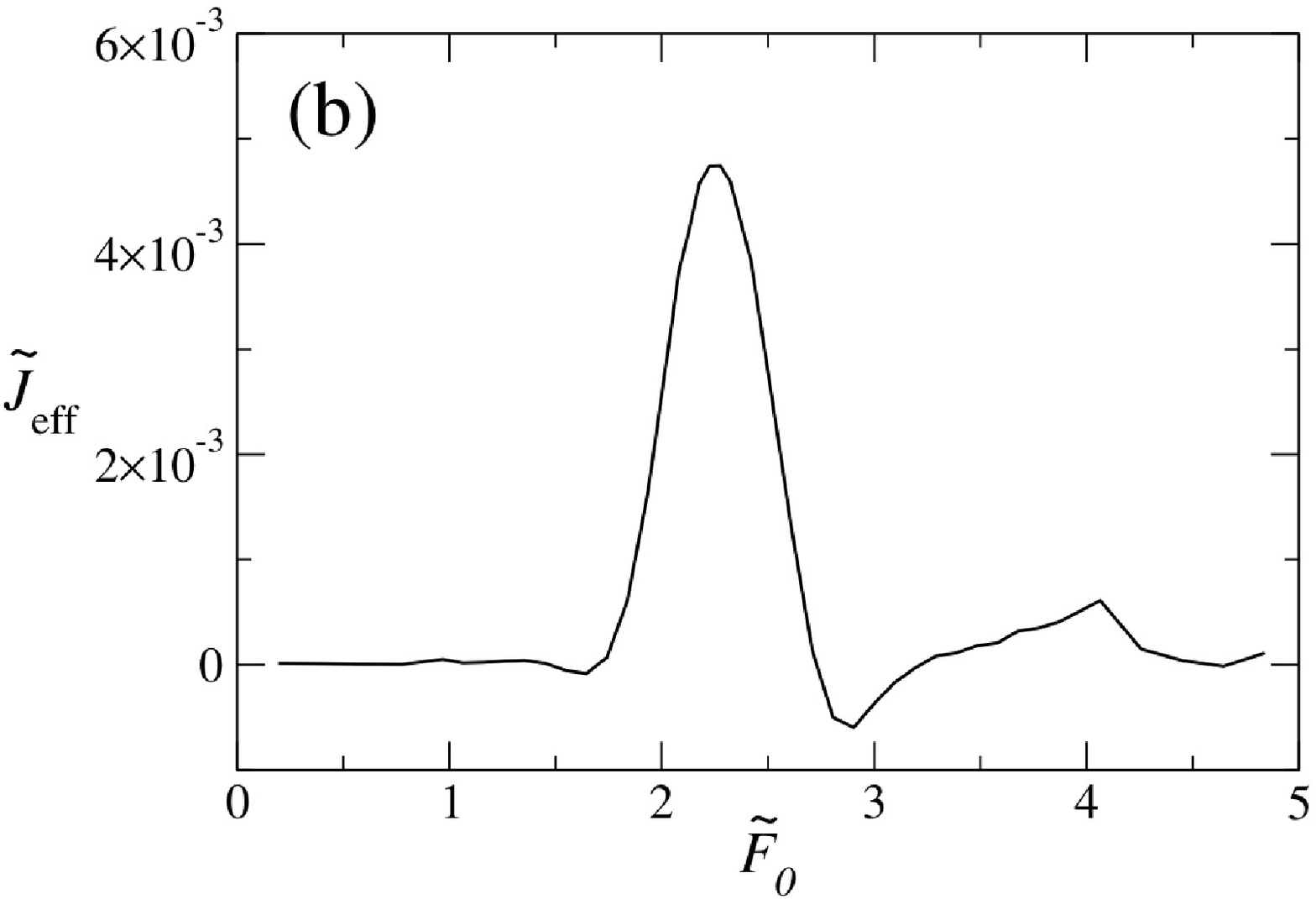}}
\end{center}
     \caption{
       (a) Particle current as a function of the period $\tilde T$ of the driving force. We display the particle current obtained from numerical simulations for the parameter values $p = 0.24$, $ \tilde F_ = 1.45$ and $ \theta = 0.0.24$. We observe that a current reversal occurs as we increase $T$. (b) Particle current as a function of the driving force amplitude $\tilde F_0$. The parameter values used to obtain this curve were chosen as $p=0.25$, $\tilde T = 2.06$ and $\tilde \theta = 0.048$. We observe that in this case the current reversal occurs as we vary the amplitude of the driving force. 
            }
\label{fig:J_reversals_T}
\end{figure}

Beyond the adiabatic limit our model exhibits a phenomenology that differs from the one observed in its adiabatic counterpart. First of all,  in analogy to rocked thermal ratchets, we found the presence of current reversals (CRs). This behavior occurs when the system is driven at high frequencies (fast tilting regime). In Fig~\ref{fig:J_reversals_T}a we can appreciate the phenomenon of CRs when we plot the particle current as a function of the period $\tilde T$ of the driving force. In Fig~\ref{fig:J_reversals_T}b we observe that the CRs are also exhibited in the system when we vary the forcing amplitude for fixed values of $\tilde T = 2.06$ and $p = 0.25$. We should emphasize that by varying continuously the period $\tilde T$, the CR is exhibited in a non-negligible window of the parameter $\tilde T \in (0.75,3.2)$.  The latter means that the phenomenon of CRs is robust under perturbations in the parameters controlling the driving force, implying that its is not necessary a fine tuning of the parameter to found current reversals in this class of systems.

%
%
%%==================== FIGURE =========================
%%
%%==================== FIGURE =========================
%
%
%
%

\begin{figure}[t]
\begin{center}
\scalebox{0.44}{\includegraphics{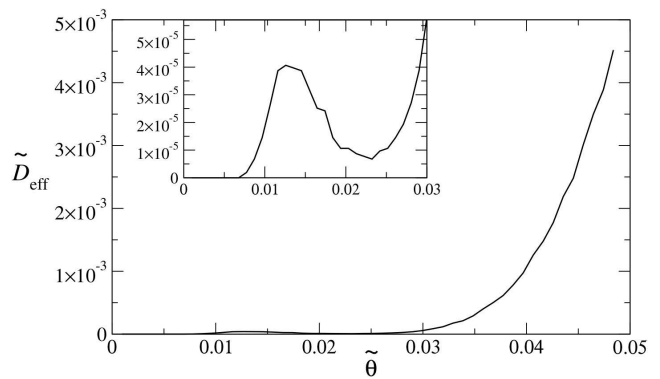}}
\end{center}
     \caption{
       Effective diffusion coefficient as a function of dimensionless temperature. We observe the estimated $\tilde D_{\mathrm{eff}}$ from numerical simulations of the stochastic differential equation~\eqref{eq:langevin} using the parameter values $p=0.25$, $\tilde F = 0.96$ and $\tilde T = 5.16$. The inset shows that the diffusion coefficient does not behaves monotonically with the temperature, which is counterintuitive phenomenon previously found in other systems (see for example Refs.~\cite{marchenko2012diffusion,marchenko2017temperature,marchenko2012anomalous,marchenko2018enhanced,rsg2014effective}). The occurrence is mainly due an interplay between the noisy and the deterministic dynamics as it has been observed in~\cite{rsg2014effective}.
       }
\label{fig:D_nonmonotonic}
\end{figure}

Another phenomenon exhibited by our model, which is worth mentioning here, is the non-monotonic dependence of the diffusion coefficient on temperature. This phenomenon is important in the sense that it implies a counterintuitive behavior of the particles diffusing in a medium. To be precise, in recent years it has been shown that, in some systems out of equilibrium, an increase in temperature does not implies always an increase in the diffusion coefficient~\cite{marchenko2012diffusion,marchenko2017temperature,marchenko2012anomalous,marchenko2018enhanced,rsg2014effective,spiechowicz2017brownian,spiechowicz2016non}. In Fig.~\ref{fig:D_nonmonotonic} we display the behavior of the diffusion coefficient $\tilde  D_{\mathrm{eff}}$ with respect to the dimensionless temperature $\tilde \theta$, maintaining fixed the amplitude $\tilde F_0 = 0.96$ and the period $\tilde T = 5.16$ of the time-dependent external forcing. We can appreciate that in a small window the diffusion coefficient does not behaves as a monotonically  increasing function of the temperature. This phenomenon has been shown to occur in a window frequencies around $\omega := 1/{\tilde T} \approx 0.193$ of size nearly $\Delta \omega \approx 0.06$. This fact suggests that this phenomenon is robust in the parameter space. The occurrence of this phenomenon can be actually a direct consequence of the fact that  it is already present in the case of a static driving force~\cite{rsg2014effective}. It has been shown that the non-monotonic behavior on temperature of the diffusion coefficient occurs in the particle-polymer model for disordered systems under the influence of a constant driving force~\cite{rsg2014effective}. In such a work, the author shows that the emergence of such a phenomenon is due to an interplay between the deterministic and noisy dynamics. Although this argument gives a simple explanation for the occurrence of the non-monotonic behavior of the diffusion coefficient it is necessary to make more exhaustive numerical and analytical studies in order to understand how does this phenomenon is affected by the parameters of the model.

%
%
%%==================== FIGURE =========================
%%
%%==================== FIGURE =========================
%
%
%
%

%$$$$$$$$$$$$$$$$$$$$$$$$$$$$$$$$$$$$$$$$$$$$$$$$$$$$$$$$$$$$$$
%$$$$$$$$$$$$$$$$$$$$$$$$$$$$$$$$$$$$$$$$$$$$$$$$$$$$$$$$$$$$$$
\section{Conclusions}
\label{sec:conclusions}
%$$$$$$$$$$$$$$$$$$$$$$$$$$$$$$$$$$$$$$$$$$$$$$$$$$$$$$$$$$$$$$
%$$$$$$$$$$$$$$$$$$$$$$$$$$$$$$$$$$$$$$$$$$$$$$$$$$$$$$$$$$$$$$

We have shown that a disordered medium produced under non-equilibrium conditions is able to rectify the motion of Brownian particles in a similar way  as it is achieved by rocket thermal ratchet systems. This rectification phenomenon is not the result of an asymmetric potential (a requirement unavoidable in ratchet systems) but the result of an interplay between noise and a coarse grained asymmetry resulting from the irreversibility of the process producing the substrate (the random potential). We have shown the occurrence of this phenomenon by taking a simple stochastic process to build up the disordered medium. The model we have chosen is a three-states Markov having a non-equilibrium steady state. The structures produced by such a Markov chain (called here ``polymers'') can be considered as out-of-equilibrium structures. By using the particle polymer model for particles moving on disordered media, we have shown that this simple system is able to rectify the motion at a finite temperature. We have also shown that at the deterministic limit, our model is unable to rectify motion due to the symmetry of the potential profiles. These two facts implies that the rectification phenomenon we report is not induced by an asymmetry of the potential profile as it occurs in deterministic ratchet systems. It is also worth mentioning that we have given an exact expression for the particle current in the adiabatic limit, based on recent works on the transport of Brownian particles on disordered media. We have also explored our model beyond the adiabatic limit by means of numerical simulations. In this regime we have found that our model exhibits current reversals as a function of both, the period and the amplitude of the external driving force. Moreover, in this system we have also reported the presence of the non-monotonic dependence of diffusion coefficient as a function of temperature. This phenomenon is important because it is a counter-intuitive behavior of Brownian particles, which is the enhancement of the dispersion by decreasing the temperature. It is also worth mentioning that, although we have used a  specific model for the Markov matrix $\mathbf{Q}$, our results can be extended to other Markov models such as those obtained by modeling out-of-equilibrium structures (see for example the work of Provata~\cite{Provata2014} where it is estimated a Markov matrix from real DNA sequences). The emergence of a net particle current will depend on the spatial irreversibility, a property which occurs whenever the entropy production of the process is positive.

In conclusion this simple model has exhibited several phenomena, already observed in other systems, but by means of a different mechanism, in which the out-of-equilibrium features of the substrate plays a central role. It would be interesting to analyze the case in which the system we propose exhibits other phenomena such as, for example, stochastic resonance~\cite{saikia2011stochastic,schiavoni2002stochastic} or the resonant response~\cite{rsg2012resonant}. We believe that our findings might contribute to a better understanding of the transport properties of mesoscopic systems where the disorder of the substrate and its non-equilibrium properties play an important role.

%$$$$$$$$$$$$$$$$$$$$$$$$$$$$$$$$$$$$$$$$$$$$$$$$$$$$$$$$$$$$$$
%$$$$$$$$$$$$$$$$$$$$$$$$$$$$$$$$$$$$$$$$$$$$$$$$$$$$$$$$$$$$$$
\section*{Acknowledgments} 
%$$$$$$$$$$$$$$$$$$$$$$$$$$$$$$$$$$$$$$$$$$$$$$$$$$$$$$$$$$$$$$
%$$$$$$$$$$$$$$$$$$$$$$$$$$$$$$$$$$$$$$$$$$$$$$$$$$$$$$$$$$$$$$

The author thanks  Cesar Maldonado for carefully reading the manuscript and his comments and suggestions that improved this work. The author also thanks the Instituto de F\'isica UASLP as well as the Instituto Potosino de Investigaci\'on Cient\'ifica y Tecnol\'ogica (IPICyT) for the warm hospitality during a sabbatical leave.

\nocite{*}

\bibliography{NoiseInducedRectification}% Produces the bibliography via BibTeX.

\end{document}